\documentclass[11pt,a4paper]{article}
\usepackage{amsmath,amssymb}
\usepackage{epsfig,graphicx}
\usepackage{array}
\usepackage{multirow}
\usepackage{hyperref}

\usepackage{tikz}
\usetikzlibrary{trees}
\usetikzlibrary{decorations.pathmorphing}
\usetikzlibrary{decorations.markings}
\usepackage{verbatim}

\usepackage{bm}

\newcommand{\be}{\begin{equation}}
\newcommand{\ee}{\end{equation}}

\newcommand{\ar}{\rightarrow}

\begin{document}

\tikzset{
    photon/.style={decorate, decoration={snake}, draw=red},
    electron/.style={draw=blue, postaction={decorate},
        decoration={markings,mark=at position .55 with {\arrow[draw=blue]{>}}}},
    gluon/.style={decorate, draw=magenta,
        decoration={coil,amplitude=4pt, segment length=5pt}}
}

\title{$\mu\ar e \gamma$ decay versus $\mu\ar eee$ bound and lepton flavor violating processes in supernova}
\author{O.~V.~Lychkovskiy$^a$\footnote{{\bf e-mail}: lychkovskiy@itep.ru},
M.~I.~Vysotsky$^{a}$
\\
$^a$ \small{\em Institute for Theoretical and Experimental Physics} \\
\small{\em 117218, B.Cheremushkinskaya 25,
Moscow, Russia}
}
\date{\today}
\maketitle

\begin{abstract}

Even tiny lepton flavor violation (LFV) due to some New Physics is able to alter the conditions inside a collapsing supernova core and probably to facilitate the explosion. LFV emerges naturally in a see-saw type II model of neutrino mass generation. Experimentally LFV beyond the Standard Model is constrained by rare lepton decay searches. In particular, strong bounds are imposed on the $\mu\ar e e e$ branching ratio and on the $\mu-e$ conversion in muonic gold. Currently the $\mu\ar e \gamma$ is under investigation in the MEG experiment which aims at dramatic increase of sensitivity in the next three years. We search for a see-saw type II LFV pattern which fits all the experimental constraints, provides ${\rm Br}(\mu\ar e \gamma)\gtrsim {\rm Br}(\mu\ar eee)$ and ensures a rate of LFV processes in supernova high enough to modify the supernova physics. These requirements are sufficient to eliminate almost all freedom in the model. In particular, they lead to a prediction $0.5 \cdot 10^{-12} \lesssim {\rm Br}(\mu\ar e \gamma) \lesssim 6 \cdot 10^{-12},$ which is testable by MEG in the nearest future. The considered scenario also constrains neutrino mass-mixing pattern and provides lower and upper bounds on $\tau$-lepton LFV decays. We also briefly discuss a model with a single bilepton in which the $\mu\ar e e e$ decay is absent at the tree level.

\end{abstract}

\section{Introduction}


Theoretical description of the collapse driven supernova explosion is an important unsolved problem in astrophysics. Modern computer simulations of the explosion have already reached high level of sophistication. Despite this they can not self-consistently explain the ejection of the supernova envelope in the whole range of relevant presupernova masses and metallicities. Usually the Standard Model is used as a microphysical input in the simulations.  However Lepton Flavor Violation (LFV) due to some New Physics at a $\sim 1$ TeV scale can substantially alter the conditions inside the collapsing core \cite{Kolb:1981mc,Fuller:1987,Amanik:2004vm,Amanik:2006ad,Lychkovskiy:2010ue}.
In particular, LFV tends to increase the neutrino luminosity thus facilitating the explosion and modifying the expected neutrino signal \cite{Lychkovskiy:2009pm,Lychkovskiy:2010ue,Lychkovskiy:2010xh}. Therefore if the true underlying theory beyond the Standard Model violates lepton flavor at a certain level, then LFV processes should be included in the supernova simulations in order to get reliable results\footnote{In the present paper we consider LFV processes {\it other than} neutrino oscillations. The latter do not occur below the neutrino sphere because of the high matter density of the supernova core. Therefore they do not affect the neutrino transport below the neutrino sphere.}.

One of the appealing SM extensions is the see-saw type II model of neutrino mass generation~\cite{Magg:1980ut}.
In our previous papers in collaboration with S. Blinnikov \cite{Lychkovskiy:2010ue,Lychkovskiy:2010xh} we have shown that under certain conditions this model predicts the rates of LFV processes in supernova high enough to alter the supernova physics. In the present paper we continue to explore the see-saw type II model.

LFV is constrained by experiments looking for rare processes with charged  leptons. By now only upper limits on corresponding transition probabilities are reported. However, a dramatic increase of statistics in such experiments is expected. In particular, MEG collaboration \cite{Mihara:2010zz} plans to reach ${\rm few} \times 10^{-13}$  sensitivity for ${\rm Br}(\mu\ar e \gamma)$ in the next few years. The {\it preliminary} result of the year 2009 run reads  ${\rm Br}(\mu\ar e \gamma) < 1.5 \cdot 10^{-11}$ at $90\%$ CL \cite{MEG ICHEP 2010}, which is already close to the best previous result due to MEGA experiment \cite{MEGA}:
\be
{\rm Br}(\mu\ar e \gamma) < 1.2 \cdot 10^{-11}, ~~~ 90\% {\rm CL.}
\ee

In the present paper we consider a scenario in which the $\mu\ar e \gamma$ decay probability is large enough to be measured by MEG in the nearest future, i.e.
\be\label{muegamma}
{\rm Br}(\mu\ar e \gamma) = x \cdot 10^{-12}
\ee
with $x$ being of order of 1. While considering this scenario one should take into account a strong experimental bound on $\mu\ar eee$ decay put by SINDRUM collaboration \cite{Bellgardt:1987du},
\be\label{mueee}
{\rm Br}(\mu\ar eee) < 1.0 \cdot 10^{-12}, ~~~ 90\% {\rm CL.}
\ee
{\it Generically} in the see-saw type II model $\mu\ar eee$ decay proceeds at tree level, while $\mu\ar e \gamma$ decay -- through one loop. Therefore  generically ${\rm Br}(\mu\ar e \gamma)\ll {\rm Br}(\mu\ar e e e)$ and the proposed above  scenario with ${\rm Br}(\mu\ar e \gamma)\sim 10^{-12}$ is ruled out. However for certain values of model parameters the $\mu\ar eee$ decay is suppressed at tree level and the considered scenario may be realized \cite{Kakizaki:2003jk,Chun:2003ej}. Is it possible to satisfy an additional requirement of sufficiently large (i.e. relevant for neutrino transport) LFV rate in supernova? The goal of the present paper is to explore this question. The result is as follows: we find a region in the parameter space of the model in which the answer is affirmative. We call this region a "Golden Domain" of the see-saw type II model. Roughly speaking this  Golden Domain corresponds to the normal neutrino mass hierarchy and $\theta_{13} > 2^o;$ in this Domain the rates of LFV processes in supernova are high enough to alter the SN physics whenever ${\rm Br}(\mu\ar e \gamma) \gtrsim    10^{-12}.$ The upper bound ${\rm Br}(\mu\ar e \gamma) <   5.8 \cdot 10^{-12}$ is derived in our model from the experimental upper bound on $\mu-e$ conversion in muonic Au atom.

The rest of the paper is organized as follows. In sect. 2 the see-saw type II model is reviewed. In sect. 3 the criterion is derived which assures that LFV processes in supernova  alter the supernova physics significantly. In sect. 4 LFV charged lepton decays and $\mu-e$ conversion are discussed. In sect. 5 interrelations between various bounds and restrictions are established and the Golden Domain of the parameter space of the see-saw type II model is presented. In sect. 6 we compare our results to what may be expected in other models, namely, in a model with a single charged bilepton and in the MSSM. In sect. 7 we summarize our results.



\section{See-saw type II.}
In the see saw type II model \cite{Magg:1980ut} a heavy scalar triplet $\bf \Delta$ is introduced which is responsible for the generation of Majorana neutrino masses. The triplet is coupled to leptons and to the SM Higgs boson, the latter coupling producing a vacuum expectation value for the neutral component of the triplet. The neutrino masses are proportional to this vev.

The see-saw type II Lagrangian contains two major ingredients,
a scalar-lepton interaction,
\be
{\cal L}_{ll\Delta}=\sum_{l,l'}\lambda_{l l'} \overline{ L_l^c} i\tau_2 \Delta L_{l'}+h.c.,
\label{llDelta}
\ee
and a scalar potential, which in its minimal form reads
\be
V= -M_H^2 H^\dagger H + f(H^\dagger H)^2+ M_\Delta^2 Tr (\Delta^\dagger\Delta) + \frac{1}{\sqrt{2}}(\tilde \mu H^T i\tau_2 \Delta^\dagger H+h.c.).
\label{Delta2}
\ee
Here
\be
\Delta\equiv {\bm \Delta \bm \tau}/\sqrt{2}=
\begin{pmatrix}
\Delta^+/\sqrt{2} & \Delta^{++}\\
\Delta^0 & -\Delta^+/\sqrt{2}
\end{pmatrix},
\ee
$L_l\equiv
\begin{pmatrix}
(\nu_{l})_L\\
l_{L}
\end{pmatrix}
$ is a doublet of left-handed leptons of flavor $l=e,\mu,\tau$,
$H$ is a Higgs doublet,
$\tilde \mu$ is a parameter with the dimension of mass.

Note that due to the anticommutation of the fermion fields $3\times3$ matrix $\Lambda\equiv||\lambda_{ll'}||$ is symmetric,
\be\label{symmetry of lambda}
\Lambda^T=\Lambda.
\ee

The vev of the neutral component of the triplet reads
\be
\langle\Delta^0\rangle=\frac{\tilde\mu v^2}{2 \sqrt{2} M_\Delta^2},
\ee
where $ v\equiv \sqrt{2} \langle H^0\rangle = 246$ GeV. Due to the triplet vev neutrinos acquire the Majorana mass according to
\be
m = 2 \langle\Delta^0\rangle \Lambda,
\label{m-lambda relation}
\ee
where $m\equiv ||m_{l l'}||$ is a neutrino mass matrix in a flavor basis. One gets that in the see-saw type II model neutrino mass matrix $m$ is proportional to the coupling matrix $\Lambda.$

The neutrino mass matrix in the flavor basis is obtained from the diagonal mass matrix through the following transformation \cite{Schechter:1980gr}:
\be\label{mass matrix}
m= U^* \cdot {\rm diag}(m_1,m_2,m_3) \cdot U^\dagger,
\ee
with $U\equiv||U_{li}||$ ($l=e,\mu,\tau,~i=1,2,3$) being a PMNS neutrino mixing matrix,
$$
 U =
\left( \begin{array}{ccc} 1 & 0 & 0 \\ 0 & c_{23}  & s_{23}
\\ 0 & -s_{23} & c_{23} \end{array} \right)
\left( \begin{array}{ccc} c_{13} & 0 & s_{13} e^{-i\delta}\\ 0 & 1 & 0 \\
-s_{13}e^{i\delta} & 0 & c_{13}
\end{array} \right)
\left( \begin{array}{ccc}   c_{12} & s_{12}  & 0
\\  -s_{12} & c_{12} & 0 \\ 0 & 0 & 1 \end{array} \right)\times
$$
\be\label{PMNS matrix}
\times{\rm diag}(e^{i\alpha_1/2}, e^{i\alpha_2/2},1).
\ee
The explicit expressions for entries of $m$  read \cite{Akeroyd:2007zv}
\be\label{m entries}
\begin{array}{lcl}
m_{ee} & = & a ~c_{13}^2  + s_{13}^2 m_3  e^{2 i \delta } \\
m_{\mu\mu} & = &  m_1 e^{-i \alpha_1} (s_{12} c_{23} + s_{13} e^{-i \delta
} c_{12} s_{23})^2+m_2 e^{-i \alpha_2} (c_{12} c_{23}-s_{13} e^{-i \delta } s_{12} s_{23})^2+m_3 c_{13}^2 s_{23}^2\\
m_{\tau\tau} & = & m_1 e^{-i \alpha_1} (s_{12} s_{23}- s_{13} e^{-i \delta } c_{12} c_{23})^2+m_2 e^{-i \alpha_2} (c_{12} s_{23}+ s_{13} e^{-i \delta }  s_{12} c_{23})^2+m_3 c_{13}^2 c_{23}^2\\
m_{e\mu} & = & c_{13} [d  s_{12} c_{12} c_{23} +s_{13} e^{i \delta }  s_{23} (m_3-a e^{-2 i \delta })]\\
m_{e\tau} & = & c_{13}[-d  s_{12} c_{12} s_{23} +  s_{13} e^{i \delta } c_{23} (m_3-a e^{-2 i \delta }) ] \\
m_{\mu\tau} & = & s_{23} c_{23}(-b+c_{13}^2  m_3) - s_{13} d e^{-i \delta}  s_{12}c_{12}(c_{23}^2-s_{23}^2) + s_{13}^2 a e^{-2i\delta} s_{23} c_{23}  \\
\end{array}
\ee
Here we define the parameters with the dimension of mass:
\be
\begin{array}{lcl}
a   &   \equiv   &  m_1 e^{-i \alpha_1} c_{12}^2+m_2 e^{-i \alpha_2} s_{12}^2,    \\
b   &   \equiv   &  m_1 e^{-i \alpha_1} s_{12}^2 + m_2 e^{-i \alpha_2}  c_{12}^2,  \\
d   &   \equiv   &  m_2e^{-i \alpha_2} -m_1 e^{-i \alpha_1}.                        \\
\end{array}
\ee

The best experimental bound on the mass of the doubly charged scalar $\Delta^{--}$ (which we are mainly interested in) is reported by the D0 collaboration \cite{D0}:
\be
M_{\Delta^{--}}>150 {\rm~GeV~~~}95\% {\rm~CL.}
\ee
A slightly weaker bound was earlier reported by the CDF collaboration \cite{CDF}. Prospects for $\Delta^{--}$ searches on LHC are discussed in a recent paper \cite{Akeroyd:2011zz}.

\section{LFV processes in supernova}

See-saw II gives rise to the following flavor changing reactions in supernova \cite{Lychkovskiy:2010ue}:\footnote{It was argued in \cite{Lychkovskiy:2010ue} that only reactions with $|\Delta L_e|,|\Delta L_\mu|,|\Delta L_\tau|=0,2$ are relevant because non-diagonal matrix elements of $\Lambda$ should be small in order to suppress yet unobserved LFV decays of charged leptons. This conclusion is valid {\it generically}; however, in the present paper we consider a special domain in the model parameter space in which the  $\mu\ar e \gamma$ decay probability is close to its experimental bound. Therefore we should consider all LFV reactions. }
\be\label{LFV processes in SN}
\begin{array}{rcll}
e^- e^- &  \rightarrow & \mu^- \mu^-, &\\
e^-\nu_e & \rightarrow &  \mu^-\nu_{e,\mu,\tau}, &\\
e^-\nu_e & \rightarrow &  e^-\nu_{\mu,\tau}, &\\
\nu_e\nu_e & \rightarrow &  \nu_l\nu_l,& l=\mu,\tau\\
\nu_e\nu_e & \rightarrow &  \nu_l\nu_{l'}, &  l,l'=e,\mu,\tau,~~l\neq l'.
\end{array}
\ee
All above processes are described by a tree diagram with $\Delta$ in $s$-channel. E.g. the first process is described by the following diagram:

\begin{figure}[h]
\centerline{\includegraphics{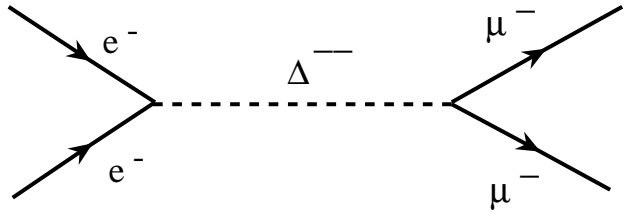}}
\caption{$ee\rightarrow \mu\mu$ LFV transition mediated by the doubly charged scalar $\Delta^{--}$ }
\label{ee->mumu}
\end{figure}
Neglecting the electron mass one gets the following cross sections:
\be\label{sigmas}
\begin{array}{lcl}
\sigma(ee\rightarrow \mu\mu) & =&
(|\lambda_{ee}|^2|\lambda_{\mu\mu}|^2/M_\Delta^4)(1-m_\mu^2/2E^2)\sqrt{1-m_\mu^2/E^2}~E^2/2\pi,
\\
\sigma(e\nu_e\rightarrow \mu\nu_l)
&=&
(|\lambda_{ee}|^2|\lambda_{\mu l}|^2/M_\Delta^4)(1-m_\mu^2/4E^2)^2E^2/2\pi, ~~~ l=e,\mu,\tau,
\\
\sigma(e\nu_e\rightarrow e\nu_l)
&=&
(|\lambda_{ee}|^2|\lambda_{e l}|^2/M_\Delta^4)E^2/2\pi, ~~~ l=\mu,\tau,
\\
\sigma(\nu_e\nu_e\rightarrow \nu_l\nu_l)
&=&2(|\lambda_{ee}|^2|\lambda_{ll}|^2/M_\Delta^4)E^2/\pi,~~~ l=\mu,\tau,
\\
\sigma(\nu_e\nu_e\rightarrow \nu_l\nu_{l'})
&=&
4(|\lambda_{ee}|^2|\lambda_{ll'}|^2/M_\Delta^4)E^2/\pi,~~~ l,l'=e,\mu,\tau,~~~l\neq l',
\end{array}
\ee
where $E$ is the energy of the initial electron or neutrino in the center of momentum frame.\footnote{These cross sections were calculated in \cite{Lychkovskiy:2010ue}; however, unfortunately, some numerical factors in \cite{Lychkovskiy:2010ue} are incorrect. Namely, $\sigma(e\nu_e\rightarrow \mu\nu_l)$ and $\sigma(\nu_e\nu_e\rightarrow \nu_l\nu_l)$ in \cite{Lychkovskiy:2010ue} have erroneous extra factors $1/2$ and  $1/4$ correspondingly.  }

The rate of conversion of electron flavor to $\mu$- and $\tau$- flavors inside the proto-neutron star can be estimated as
\be
R_{\rm LFV} \simeq \frac{n_e^2}{2} \sigma(e e \rightarrow \mu\mu) + n_e n_{\nu_e} \sum_{f,f'} \sigma(e\nu_e\rightarrow f f') +\frac{n_{\nu_e}^2}{2} \sum_{f,f'}  \sigma(\nu_e\nu_e\rightarrow ff'),
\ee
where $f$ and $f'$ denote various final neutrinos and charged leptons, see eq.(\ref{LFV processes in SN}). If this rate is comparable with the rate of decrease of the total lepton number due to neutrino diffusion out of the proto-neutron star, $R_{\rm diff},$ then the physics of the collapse is substantially altered compared to the SM case. In particular, the neutrino signal is modified and the  explosion is probably facilitated \cite{Lychkovskiy:2010ue,Lychkovskiy:2010xh}. To be specific, we demand that
\be\label{SN bound}
R_{\rm LFV}>R_{\rm diff}\simeq 4\cdot 10^{36}{\rm cm}^{-3}{\rm s}^{-1}.
\ee
The latter numerical value is based on the supernova simulations from ref.\cite{Burrows:1986me}. 
Matter in the center of supernova after core bounce is characterized by $n_B\simeq 2\cdot 10^{38}$ cm$^{-3},$ $Y_e\equiv n_e/n_B \simeq 0.28,$ $Y_{\nu_e}\equiv n_{\nu_e}/n_B \simeq 0.07,$ $\mu_e\simeq (240-280)$MeV, $\mu_{\nu_e}\simeq (160-220)$MeV (one can get these values e.g. from paper \cite{Burrows:1986me} or using the open-code programm BOOM described in \cite{Boom}).
For the numerical estimates we conservatively take $E=160$MeV. We use the above numerical values to establish the relation between the $\mu\rightarrow e\gamma$ decay probability, $R_{\rm LFV}$ and $R_{\rm diff}$ in sect. 5.


\section{Rare lepton decays}

\begin{table}[t]
\vspace{0.2cm}
\begin{center}
\begin{tabular}{c|c|c|c|c}
process & experimental & \multicolumn{3}{|c} {Br(process)$/x$} \\
\cline{3-5}
 & upper bound on Br & I & II & III \\
\hline
$\mu  \rightarrow  e\gamma$  & $1.2 \cdot 10^{-11}$ & \multicolumn{3}{|c} {$ 10^{-12} $} \\
\hline
$\mu^-  \rightarrow  e^+e^-e^-$   & $1.0\cdot 10^{-12}$ &  $\lesssim 10^{-13}$ & $4.1\cdot 10^{-13}$ & $ 3.3 \cdot 10^{-13}$  \\
\hline
\hline
$\mu ~{\rm Au} \rightarrow  e~ {\rm Au}$ ($M_\Delta=150$~GeV)& \multirow{2}{*}{$7 \cdot 10^{-13}$} & $1.2 \cdot 10^{-13}$ & $1.9 \cdot 10^{-13}$ & $1.7 \cdot 10^{-13}$ \\
\cline{1-1}
\cline{3-5}
$\mu ~{\rm Au} \rightarrow  e~ {\rm Au}$ ($M_\Delta=1$~TeV) & & $3.1 \cdot 10^{-13}$ & $ 4.2 \cdot 10^{-13}$ & $3.8 \cdot 10^{-13}$ \\
\hline
\hline
$\tau^-  \rightarrow  \mu^+\mu^-\mu^-$ & $3.2 \cdot 10^{-8}$  & $ 1.0 \cdot 10^{-9}$ & $ 3.4 \cdot 10^{-10}$ & $ 9.1 \cdot 10^{-10}$\\
\hline
$\tau^-  \rightarrow  e^+\mu^-\mu^-$  & $2.3 \cdot 10^{-8}$  &  $ 7.6 \cdot 10^{-11}$  &  $ 3.9 \cdot 10^{-11}$  &  $ 6.1 \cdot 10^{-11}$\\
\hline
$\tau^-  \rightarrow  e^+e^-e^-$  & $3.6 \cdot 10^{-8}$  &  $ 9.6 \cdot 10^{-13}$  &  $ 9.3 \cdot 10^{-13}$  &  $ 6.7 \cdot 10^{-13}$\\
\hline
$\tau^-  \rightarrow  \mu^+e^-e^-$  & $2.0 \cdot 10^{-8}$  & $ 1.3 \cdot 10^{-11}$  & $ 8.4 \cdot 10^{-12}$   & $ 1.0 \cdot 10^{-11}$   \\
\hline
$\tau^-  \rightarrow  e^+e^-\mu^-$  & $ 2.7 \cdot 10^{-8}$  & $ \lesssim 10^{-11}$  & $ 3.8 \cdot 10^{-13}$   & $  4.2 \cdot 10^{-13}$   \\
\hline
$\tau^-  \rightarrow  \mu^+e^-\mu^-$  & $3.7 \cdot 10^{-8}$  & $  \lesssim 10^{-13}$  & $ 3.4 \cdot 10^{-12}$   & $ 6.3 \cdot 10^{-12}$   \\
\hline
\hline
$\tau  \rightarrow  \mu\gamma$ & $3.3 \cdot 10^{-8}$  & $ 1.6 \cdot 10^{-11}$  & $ 5.4 \cdot 10^{-12}$   & $ 1.4 \cdot 10^{-11}$   \\
\hline
$\tau  \rightarrow  e\gamma$ & $4.4 \cdot 10^{-8}$   & $ 3.5 \cdot 10^{-13}$   & $ 2.7 \cdot 10^{-13}$   & $  3.0 \cdot 10^{-13}$  \\
\hline
%
\end{tabular}
\end{center}
\caption{ Lepton flavor violating processes: experimental constraints \cite{Nakamura:2010zzi,Bertl:2006up} and predicted values. The latter correspond to the three selected points from the Golden Domain of the see-saw type II model (see Table 2) and are normalized to the $x\equiv {\rm Br}(\mu  \rightarrow  e\gamma)/10^{-12}.$ A "$\lesssim$" sign is used whenever the probability of the process vanishes at tree level and is given by higher-order loop diagrams (see text for further details). The "branching ratio"~ for the $\mu  -  e$ conversion on Au is defined as ${\rm \Gamma}_{\rm Au}{\rm(} \mu  \ar e {\rm )}/{\rm \Gamma}_{\rm capt},$ where ${\rm \Gamma}_{\rm capt}=13.07\cdot10^{6}$ s$^{-1}$ \cite{Suzuki:1987jf} is the muon capture rate in the muonic gold.
\label{table rare processes}
			}
\end{table}

Present experimental constraints on so-called "rare"~(in fact yet unobserved) LFV lepton processes are summarized in the second column of Table \ref{table rare processes}. A detailed analysis of LFV charged lepton decays mediated by scalar triplet is given in \cite{Akeroyd:2009nu}. Three-lepton rare decays normally proceed at the tree level and their widths are given by
\be\label{mu-3e width}
{\rm \Gamma(} \mu^-\ar e^+e^-e^- {\rm )} = \frac{m_\mu^5}{768 \pi^3 M_\Delta^4}|\lambda_{e\mu}\lambda_{ee}|^2,
\ee
\be\label{tau-l2l' width}
{\rm \Gamma(} \tau^-\ar l^+l'^-l'^- {\rm )} = \frac{m_\tau^5}{768 \pi^3 M_\Delta^4}|\lambda_{l\tau}\lambda_{l'l'}|^2,
\ee
\be\label{tau-ll'l'' width}
{\rm \Gamma(} \tau^-\ar l^+l'^- l''^- {\rm )} = \frac{m_\tau^5}{384 \pi^3 M_\Delta^4}|\lambda_{l\tau}\lambda_{l'l''}|^2~~~{\rm for}~~~l'\neq  l''.
\ee
Note that the decays with two identical leptons of equal sign in the final state, see eqs. (\ref{mu-3e width}) and (\ref{tau-l2l' width}), have an additional factor $1/2,$ compared to the decay (\ref{tau-ll'l'' width}) with different leptons of equal sign in the final state.

Radiative $l \ar l' \gamma $ decays are described by penguin diagrams, therefore  their widths contain an additional
factor $\sim\alpha$ \cite{Kakizaki:2003jk}:
\be\label{radiative decay}
{\rm \Gamma(} l \ar l' \gamma  {\rm )} = \frac{27}{16}\frac{\alpha}{4\pi}
\frac{m_l^5}{192 \pi^3 M_\Delta^4}|\lambda_{le}\lambda_{el'}^*+\lambda_{l\mu}\lambda_{\mu l'}^*+\lambda_{l\tau}\lambda_{\tau l'}^*|^2.
\ee
One could thus expect that generically  ${\rm Br} (l_1 \ar l_2 \gamma)\ll{\rm Br} (l_1 \ar l_2 l_3 l_4).$   This relation implies ${\rm Br(}\mu  \rightarrow  e\gamma {\rm)} \ll 10^{-12}$ due to the strong $\mu\ar eee$ experimental bound, which makes the $\mu  \rightarrow  e\gamma$ decay unobservable in the MEG experiment.  However, due to the fact that matrix $\Lambda$ is related to the neutrino masses and mixing, one can expect a hierarchy of couplings and therefore of decay rates. Indeed, in the next section it is shown that the above-mentioned  controversary may be avoided for a certain (allowed by the experimental data) choice of $\Lambda$. It is clear that this choice should lead to the suppression of the tree amplitude of $\mu\ar eee$ decay.

Another strong bound on LFV is imposed by the results of SINDRUM II collaboration on $\mu-e$ conversion on  gold \cite{SINDRUM II}. This experiment investigated the fate of muonic atoms with heavy nuclei. The most probable event is the capture of muon by nucleus with muon neutrino emission. A LFV mode is the $\mu-e$ transition which results in a monoenergetic electron emission. This process was first theoretically explored in ref. \cite{Weinberg:1959zz}. An approximate expression for the width of the $\mu-e$ conversion may be written in a model-independent way as follows \cite{Kakizaki:2003jk}:
\be\label{e-mu conversion}
{\rm \Gamma}_{(A,Z)}{\rm(} \mu  \ar e {\rm )} = 4\alpha^5 m_\mu^5 Z_{\rm eff}^4 Z |F_p(q^2)|^2(|A_1^L+A_2^R|^2+|A_1^R+A_2^L|^2).
\ee
Here $Z_{\rm eff}$ is an effective charge as felt by a muon bound in atom, $F_p(q^2)$ is a form factor related to the proton density in nucleus, $q^2\simeq m_\mu^2$  and $A_{1,2}^{L,R}$ are the model-dependent form factors which enter the effective low-energy LFV violating electromagnetic current,
\be
j^\alpha=\overline e [q^2\gamma_\alpha(A_1^L P_L+A_1^R P_R)+m_\mu i\sigma_{\alpha\beta}q^\beta(A_2^L P_L+A_2^R P_R)] \mu.
\ee
Formula (\ref{e-mu conversion}) demonstrates how the $\mu-e$ conversion rate depends on the quantities involved; however it strongly depends on the quantities  $Z_{\rm eff}$ and $F_p(q^2)$ which can not be expressed analytically. A thorough analysis of the $\mu-e$ transition rate is presented in refs. \cite{Czarnecki:1998iz,Kitano:2002mt}. We use their results, which are reproduced if one takes for gold $Z_{\rm eff}=33.5,$ $F_p(q^2)=0.16$ \cite{Kitano:2002mt}.

The  form-factors $A_1^L,~A_2^R$ for the see-saw type II model read \cite{Kakizaki:2003jk}
\be
A_1^L=\sum_l f_l \frac{\lambda_{el}^*\lambda_{l\mu}}{12\pi^2M_\Delta^2},~~~
A_2^R=\sum_l  \frac{3 \lambda_{el}^*\lambda_{l\mu}}{32\pi^2M_\Delta^2},
\ee
while $A_1^R$ and $A_2^L$ vanish due to the electron chirality conservation.
Here
\be
f_l = \ln\frac{m_l^2}{M_\Delta^2}+4\frac{m_l^2}{|q^2|}
+(1-2\frac{m_l^2}{|q^2|})
\sqrt{1+4\frac{m_l^2}{|q^2|}}
\ln\frac{\sqrt{|q^2|+4m_l^2}+\sqrt{|q^2|}}{\sqrt{|q^2|+4m_l^2}-\sqrt{|q^2|}},
\ee
where $|q^2|\simeq m_\mu^2.$ This general expression is simplified for given flavors:
\be
\begin{array}{l}
f_e   \simeq \ln\frac{|q^2|}{M_\Delta^2}=-18.3,\\
f_\mu \simeq \ln\frac{m_\mu^2}{M_\Delta^2}+(4-\sqrt{5}\ln\frac{3+\sqrt{5}}{2})=-16.5,\\
f_\tau\simeq \ln\frac{m_\tau^2}{M_\Delta^2}+\frac53=-11.0,\\
\end{array}
\ee
where the numerical values are given for $M_\Delta=1$ TeV.
Large logarithmic factors in $f_l$ appear due to the diagram in which the photon couples to a charged fermion in the loop. Contracting in this diagram the propagator of $\Delta$-boson one obtains the photon polarization operator which contains this famous logarithm responsible for the running of electromagnetic coupling $\alpha.$ Due to the large logarithmic factor $A_1^L$ dominates over $A_2^R$ in the probability of the $\mu-e$ conversion.

Note that all rare decay probabilities have the same $\sim M_\Delta^{-4}$ dependence on the scalar mass. If one fixes the coupling matrix $\Lambda$  up to a common factor $\lambda$ and introduces an effective four-fermion constant $G_{\rm LFV}=\lambda^2/M_\Delta^2,$ then all rare decay probabilities will depend only on $G_{\rm LFV}$ but not on $M_\Delta$ and $\lambda$ by themselves. For this reason the values of the rare decay widths in the third column of Table 1 do not explicitly depend  on $M_\Delta.$ In contrast, the $\mu-e$ conversion probability has an additional logarithmic dependence on  $M_\Delta,$ therefore we quote two different values for it in Table 1 which correspond to two different values of $M_\Delta$ (our reference value $M_\Delta=1$ TeV and the experimental lower bound $M_\Delta=150$ GeV).


\section{Golden Domain of the see-saw type II model}

Now we are in a position to look for a "Golden Domain"~ of the see-saw type II model in which:
\begin{enumerate}
\item all the experimental constraints from neutrino oscillations, $\mu-e$ conversion and rare lepton decays are satisfied,
\item ${\rm Br}(\mu\ar e \gamma) \sim 10^{-12}$
(as explained above, this implies the suppression of the tree level amplitude for the $\mu\ar eee$ decay),
\item the rate of LFV in supernova is high enough to affect the neutrino transport (see eq.(\ref{SN bound})).
\end{enumerate}
A natural and convenient way to parameterize (up to an overall factor) the coupling matrix $\Lambda$ of the see-saw type II model is to use the neutrino masses,  mixing angles and phases as parameters. In this natural parametrization there are five continuous and two discrete parameters which are  not fixed (but possibly restricted) by  neutrino oscillation experiments. They are: the absolute scale of neutrino masses, angle $\theta_{13},$ phases $\delta,~\alpha_1,~\alpha_2$ (continuous), mass hierarchy, sign of $\tan\theta_{23}$ (discrete). In what follows we use this natural parametrization to explore the experimentally allowed part of the parameter space of the see-saw type II model.

To suppress the $\mu \ar eee$ decay one should choose \cite{Kakizaki:2003jk,Chun:2003ej}
\be\label{lambda_emu}
\lambda_{e \mu} \simeq 0.
\ee
Another possible way to suppress the $\mu \ar eee$ decay would be to set $\lambda_{ee} \simeq 0;$ however this would suppress the LFV processes in supernova~(\ref{LFV processes in SN}) as well -- thus, it is unappropriate.

Condition (\ref{lambda_emu}) implies that $m_{e \mu}$ should vanish. To get an idea how this can occur let us consider a case when $m_1\ll m_2\ll m_3.$  Then from eq. (\ref{m entries}) one gets
\be\label{m_emu}
m_{e\mu}  \simeq  e^{-i \alpha_2 } \cos\theta_{23}(  \frac12 \sin 2\theta_{12} m_2+\tan\theta_{23} \sin\theta_{13} e^{i (\delta+\alpha_2) }   m_3).
\ee
Remind that by now the  $\theta_{23}$ octant ambiguity persists, i.e. the values $\tan\theta_{23}\simeq \pm 1$ are experimentally allowed.
If $\tan\theta_{23}\simeq - 1$ and if $(\delta+\alpha_2)\mod 2\pi \simeq 0,$ then  the cancelation of two terms in eq. (\ref{m_emu})  occurs for $\theta_{13}\simeq 5^o$ \cite{Kakizaki:2003jk}.\footnote{The sign prescriptions in definition of matrix elements of the PMNS matrix $U$ in ref. \cite{Kakizaki:2003jk}  differ from ours. The difference is not physical; it is eliminated if one makes the  transformation $L_\mu\ar -L_\mu,~\mu_R\ar -\mu_R.$} Thus we are able to fit the condition (\ref{lambda_emu}) choosing experimentally allowed mass-mixing pattern.

\begin{figure}[t]
\centerline{
$
\begin{array}{cc}
\includegraphics[width=0.5\textwidth]{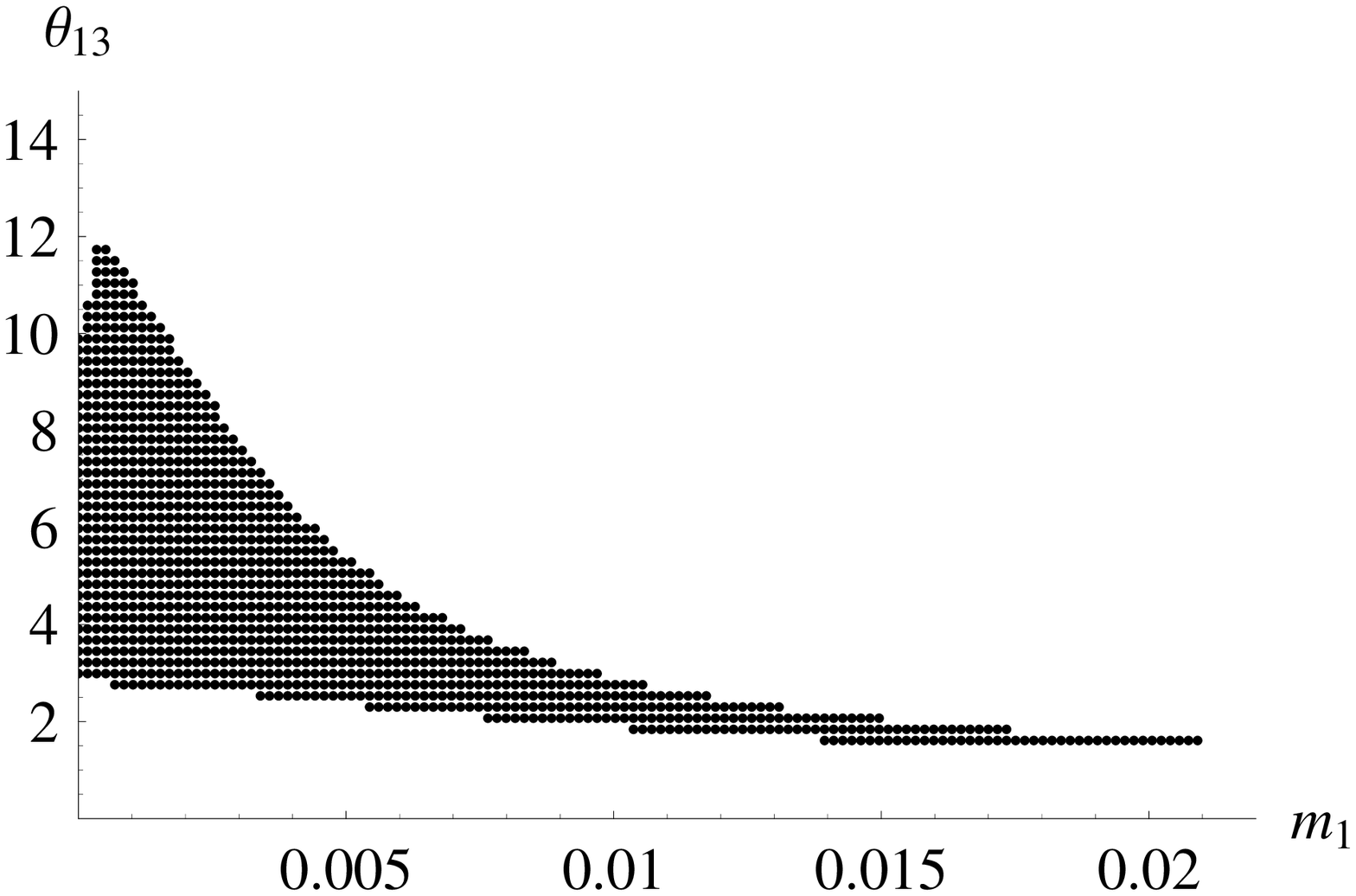} &
\includegraphics[width=0.5\textwidth]{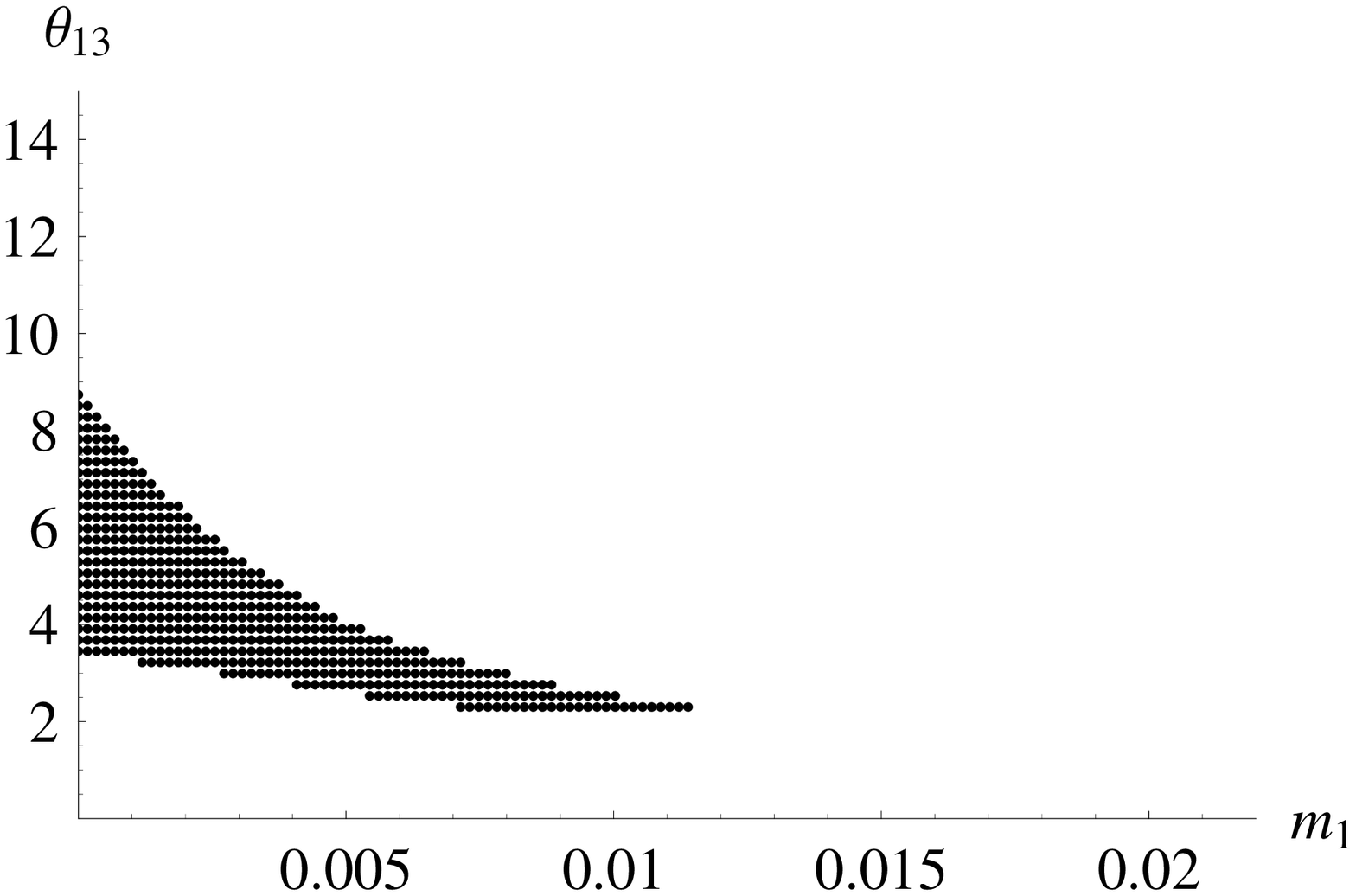}\\
\includegraphics[width=0.5\textwidth]{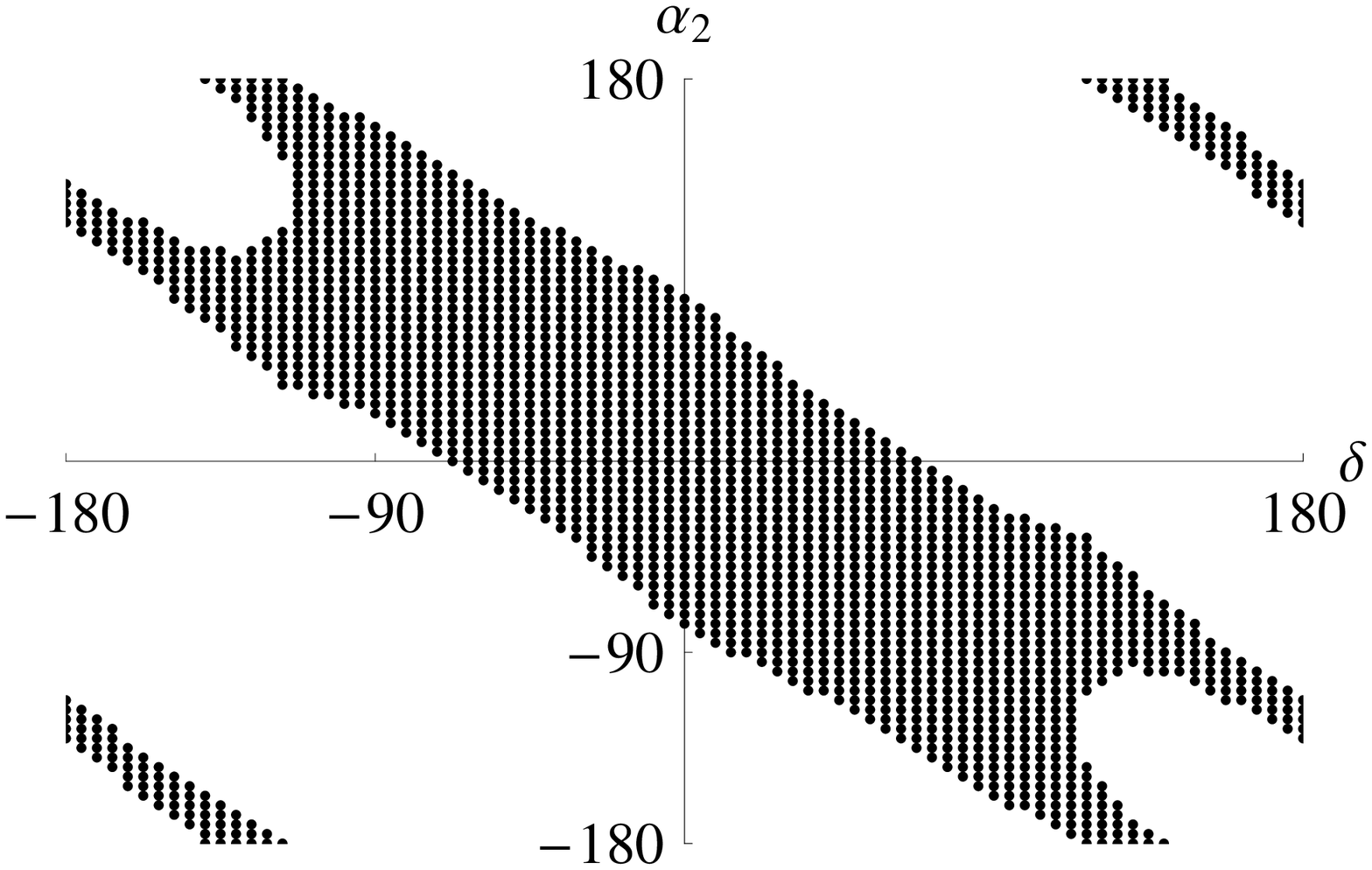} &
\includegraphics[width=0.5\textwidth]{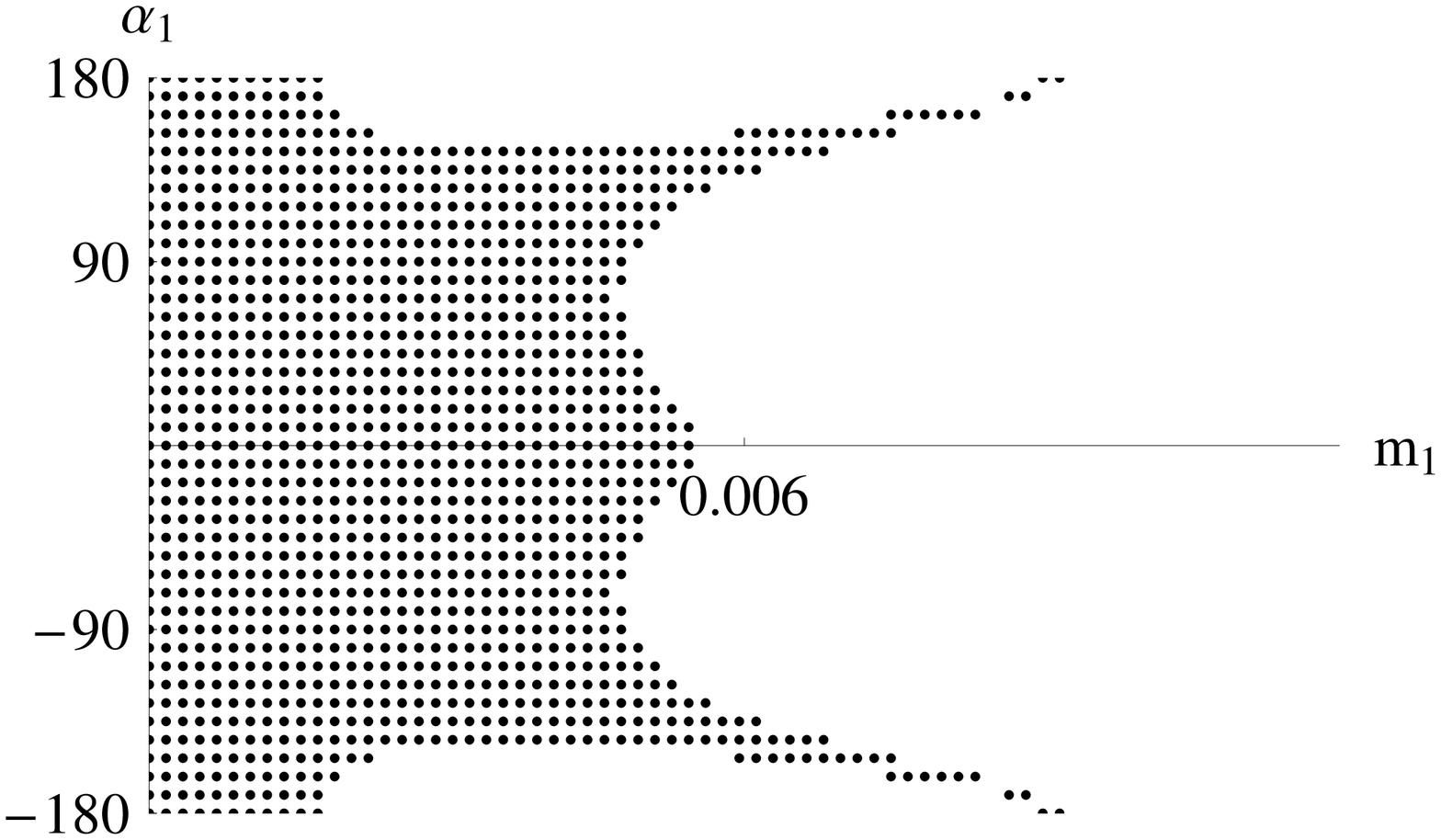}
\end{array}
$
}
\caption{\label{plot Golden Domain} Projections of Golden Domain  of the see-saw type II model. The Golden Domain consists of the points in the neutrino mass-mixing parameter space which provide ${\rm Br}(\mu\ar e \gamma) = x \cdot 10^{-12}$ and fit all the experimental constraints in the framework of the see-saw type II model. $x=3$ for the upper right plot and $x=1$ for other three plots. For all plots $\theta_{23}=135^o$ and the mass hierarchy is  normal, the masses $m_2$ and $m_3$ being related to $m_1$ through the well-known mass-squared differences, $\Delta m^2_{21}=0.76\cdot10^{-4}$ eV$^2$ and $\Delta m^2_{31}=24.3\cdot10^{-4}$ eV$^2$ \cite{PDG 2}. The remaining parameters are fixed as follows: two upper plots correspond to $\delta=\alpha_1=\alpha_2=0,$ lower left plot -- to $m_1=0,~\theta_{13}=5^o,~\alpha_1=0,$ lower right plot -- to $\theta_{13}=5^o,~\delta=\alpha_2=0.$ Masses are given in eV and angles -- in degrees.}
\end{figure}

In order to get a general picture we scan numerically the parameter space of the see-saw type II model. As a result we find a single Golden Domain of the parameter space which satisfies all the imposed requirements. Some of the 2D projections of this domain are presented on Fig. 2.
The main features of this domain are as follows.
\begin{itemize}
\item $\theta_{23} \simeq 135^o$ (second octant).
\item Normal mass hierarchy with $m_1 < m_2 \ll m_3.$ Neutrino masses can take the following values:
    \be
    0< m_1 \lesssim 0.021 ~{\rm eV,}~~~ 0.009~{\rm eV}\lesssim m_2 \lesssim 0.023 ~{\rm eV,}~~~ m_3 \simeq 0.05 ~{\rm eV}.
    \ee Moreover, as follows from Fig. 2,
     the case of quasidegenerate $m_1$ and $m_2$ (with $m_1\gtrsim 0.005$) is only marginally allowed; on the contrary,  substantially hierarchical values $m_1\ll m_2$ occupy the major part of the Golden Domain.
\item The value of $\theta_{13}$ may vary in a broad range, but can not be too small:
\be\label{theta13 bounds}
2^o \lesssim \theta_{13} \lesssim 12^o.
\ee
\item Phases $\delta$ and $\alpha_2$ should satisfy
\be
|(\delta+\alpha_2)\mod 2 \pi| \lesssim 40^o.
\ee
\item The value of $\alpha_1$ may vary in a broad range, especially when $m_1\ll m_2.$ This is easy to understand as $\alpha_1$ enters the mixing matrix only in the expression $m_1 e^{i\alpha_1/2},$ which may be disregarded when $m_1$ vanishes.
\end{itemize}

In Table 1 we show the predictions for the probabilities of LFV processes for three selected points in the parameter space. These points are defined in Table 2, and the corresponding coupling matrices are given in Table 3. Finally, the rates of LFV processes in supernova are $1.4 x R_{\rm diff},~1.8 x R_{\rm diff}$ and $~2 x R_{\rm diff}$ for points I, II and III correspondingly.

%
\begin{table}[t]
\begin{center}
$\begin{array}{|c|c|c|c|c|c|c|c|c|}
\hline
 & m_1 & m_2 &  m_3 & \theta_{12} &  \theta_{23} & \theta_{13} & \delta & \alpha_1,\alpha_2 \\
\hline
{\rm I} & 0&
0.9\cdot 10^{-2} {\rm eV}&
5 \cdot 10^{-2} {\rm eV}&
34^o &
135^o &
5^o & 0 & 0\\
\hline
{\rm II} & 0.1 \cdot 10^{-2} {\rm eV}&
0.9\cdot 10^{-2} {\rm eV}&
5 \cdot 10^{-2} {\rm eV}&
34^o &
135^o &
8^o & 0 & 0\\
\hline{\rm III} & 0&
0.9\cdot 10^{-2} {\rm eV}&
5 \cdot 10^{-2} {\rm eV}&
34^o &
135^o &
5^o & 30^o & 0\\
\hline
\end{array}$
\caption{\label{table neutrino parameters} Neutrino mass-mixing parameters for three selected points in  in the Golden Domain. These reference points are used in Table 1.
 }
\end{center}
\end{table}
%

\begin{table}[t]
\begin{center}
$\begin{array}{|c|c|c|}
\hline
{\rm I} & {\rm II} &{\rm III} \\
\hline
\left(
\begin{array}{ccc}
0.0053 & 0 & -0.099\\
0 & 0.048 & -0.037\\
-0.099 & -0.037 & 0.047
\end{array}
\right)
&
\left(
\begin{array}{ccc}
0.0057 & 0.0026 & -0.0093\\
0.0026  & 0.037 & -0.028\\
-0.0093 & -0.028 & 0.036
\end{array}
\right)
&
\left(
\begin{array}{ccc}
0.0054 & 0 & -0.01\\
0 & 0.048 & -0.037\\
-0.01 & -0.037 & 0.047
\end{array}
\right)\\
\hline
\end{array}$
\caption{\label{table coupling matrices} Coupling matrices corresponding to three selected points in the Golden Domain (see Table 2). Each matrix should be multiplied by $x^{\frac14}M_\Delta/(1~{\rm TeV}).$
        }
\end{center}
\end{table}

One can see that as soon as the $\mu\ar eee$ experimental constraint is made harmless, the most pressing current experimental bound stems from the SINDRUM II experiment on $\mu-e$ conversion, which bounds ${\rm Br}(\mu\ar e \gamma)$ from above. On the other hand, the condition (\ref{SN bound}) leads to the lower bound on ${\rm Br}(\mu\ar e \gamma).$ As a results, one obtains
\be\label{x window}
0.5\lesssim x \lesssim 6.
\ee


Note that for vanishing $\lambda_{e \mu}$ the $\mu  \rightarrow  e\gamma$  decay and $\mu  - e$ conversion on nuclei proceed only through the virtual $\tau$ or $\nu_\tau$ in the loop. Also note that in this case the tree contributions for the $\mu \ar eee,$ $\tau^- \ar e^+e^-\mu^-$ and $\tau^- \ar \mu^+e^-\mu^-$ decays vanish, and the decays proceed through the exchange of virtual photon. The width of the $\mu^- \ar e^-\gamma^*\ar  e^-e^+e^-$ process, compared to the $\mu\ar e\gamma$  decay, is suppressed by $\alpha$ and by the ratio $\Phi_3/\Phi_2 \sim1/(4\pi)$ of three-particle to two-particle phase volumes, but enhanced by a square of large logarithm $(\ln\frac{m_\tau^2}{M_\Delta^2})^2:$
\be
{\rm Br}{\rm(} \mu  \ar eee {\rm )} \sim \frac{\alpha}{4\pi} \cdot (\ln\frac{m_\tau^2}{M_\Delta^2})^2 \cdot  {\rm Br}{\rm(} \mu  \ar e\gamma {\rm )} \lesssim 10^{-1}{\rm Br}{\rm(} \mu  \ar e\gamma {\rm )}.
\ee
Analogous estimates are valid for the $\tau^- \ar e^+e^-\mu^-$ and $\tau^- \ar \mu^+e^-\mu^-$ decay probabilities. We use these estimates in Table 1.

From Table 3 it follows that $\lambda_{ll'}/M_\Delta{\rm (TeV)}<0.05.$ This allows to estimate the contribution of new scalar field to the anomalous muon magnetic moment: $\delta a \sim m_\mu^2 A_2^R<2\cdot10^{-13}.$ This is well beyond the present experimental sensitivity.

\section{Comparison with other models}

\subsection{Singlet bilepton model}

As is clear from the above discussion, the strong experimental bounds on the $\mu\ar eee$ decay and $\mu-e$ conversion on Au create a certain pressure on the allowed range of probability of the $\mu\ar e\gamma$ decay in the see-saw type II model.
It is interesting to note that there exists a ''close relative'' of the see-saw type II model in which this pressure is completely absent. This is a simple model which extends the Standard Model by one charged heavy bilepton  (i.e. scalar whith lepton number 2) which is coupled to leptons as follows:
\be\label{ll tildeDelta}
{\cal L}_{ll\tilde\Delta}=\sum_{l,l'}\tilde\lambda_{l l'} \overline{ L_l^c} i\tau_2 L_{l'}\tilde\Delta+h.c.
\ee
The difference from eq. (\ref{llDelta}) is that $\tilde\Delta$ is a singlet while $\Delta$ is a triplet (see \cite{Cuypers:1996ia} for a systematical classification of bileptons). The important feature of the above coupling is that coupling matrix $||\tilde\lambda_{l l'}||$ is antisymmetric, in contrast to a symmetric coupling matrix in the see-saw type II model. As a consequence, the $\mu\ar eee$ decay is forbidden at the tree level (as well as the decays of $\tau$-lepton with two identical leptons in the final state). As for the $\mu-e$ conversion, its probability does not obtain a large $\ln^2$ enhancement because only neutrinos (not charged leptons) enter the loop of the corresponding penguin diagram.  At the same time the $\mu\ar e\gamma$ decay probability is of the same order as in the see-saw type II model. The LFV processes in supernova can not proceed at tree level because the corresponding tree amplitudes would be proportional to $\lambda_{ee}.$ However, $e$ or $\nu_e$ may change flavor while scattering on a charged particle through the exchange of virtual photon in  $t$-channel.  An example of such a process is
\be
\nu_e p \ar \nu_{\mu} p.
\ee
Note that in case of neutrino scattering a charged lepton enters the loop of the penguin diagram, and the cross section receives the $\ln^2$ enhancement. A detailed study of LFV in this singlet bilepton model will be carried out elsewhere.

\subsection{MSSM}

In the MSSM all LFV processes proceed through the loop diagrams. The radiative decays proceed through penguin diagrams, while the three-lepton decays of $\mu$ and $\tau$ -- through   the box diagrams and (for some decays) through the penguin diagrams with a virtual photon decaying into the lepton-antilepton pair. Therefore generically ${\rm Br}(\mu\ar eee) \sim g^2 {\rm Br}(\mu\ar e\gamma).$ Moreover, heavy sleptons (not light charged leptons) enter loop diagrams, therefore there is no  $\ln^2$ enhancement of the $\mu-e$ conversion probability. Thus the above-mentioned pressure of strong experimental bounds on the $\mu\ar eee$ and $\mu-e$ conversion probabilities is absent in the MSSM. However, the absence of the tree-level LFV processes and of logarithmic enhancement of $\gamma^*$ emission amplitude generically severely suppresses the LFV rate in supernova.

Note that in the  MSSM the vertices in the above-mentioned penguin and box diagrams contain the elements of the unitary PMNS matrix $U.$ Therefore, if all sleptons are degenerate, then due to the GIM mechanism all LFV probabilities are zero. In contrast, in the above discussed  models with bileptons $||\lambda_{ll'}||$ and $||\tilde\lambda_{ll'}||$  are not unitary matrices  and therefore GIM mechanism does not work.

The degeneracy of sleptons is removed by the heavy $\tau$-lepton. For this reason amplitudes of LFV processes in the MSSM are proportional to $\sin\theta_{13}$ (see e.g. a recent work \cite{Davidkov:2011kv} and references therein). In the Golden Domain of the see-saw type II model $\theta_{13}$ also can not be too small, see eq. (\ref{theta13 bounds}), but this similarity between the see-saw type II and MSSM is accidental.

\section{Summary and conclusions}

We have discussed a number of requirements to the lepton flavor violation in the see-saw type II model. Apart from the mandatory requirement of satisfying all the experimental bounds, we impose two supplementary requirements which severely constraint parameter space of the model. The first one was previously discussed in the literature \cite{Kakizaki:2003jk,Chun:2003ej}: the branching ratio of the $\mu\ar e \gamma$ decay is of order of $10^{-12}$ (which ensures soon discovery at MEG), i.e.  close to the experimental upper bound on  the branching  ratio of the $\mu\ar eee$ decay. This is possible only if the tree amplitude of $\mu\ar eee$ decay is suppressed by a vanishing (or very small) coupling constant, either $\lambda_{ee}$ or $\lambda_{e\mu}$. The second requirement is that the rates of LFV processes in supernova are high enough to alter the supernova physics (such alteration may facilitate the explosion). This is possible in some region of the parameter space \cite{Lychkovskiy:2010ue,Lychkovskiy:2010xh}, in particular when  $\lambda_{e\mu}\simeq 0,$ but {\it not} when $\lambda_{ee}\simeq 0.$ As a consequence of the imposed requirements, we obtain a "Golden Domain" in the neutrino mass-mixing parameter space.

If one stands in the Golden Domain, the experimental results on $\mu-e$ conversion on gold \cite{SINDRUM II} impose the most  restrictive {\it upper} bound on ${\rm Br}(\mu\ar e \gamma),$ as is clear from Table 1.  On the other hand, the condition (\ref{SN bound}) on the LFV rates in supernova provides a {\it lower} bound. In total, the imposed constraints appear to be strong enough to force ${\rm Br}(\mu\ar e \gamma)$ to lie in a narrow window,
\be\label{muegamma window}
0.5 \cdot 10^{-12} \lesssim {\rm Br}(\mu\ar e \gamma) \lesssim 6 \cdot 10^{-12},
\ee
see eq. (\ref{x window}).
One should take into account that the upper bound corresponds to the minimal experimentally allowed scalar mass of $150$GeV; this bound becomes tighter if the mass is increased.

The branching ratios of the LFV $\tau$ decays in the Golden Domain of the see-saw type II model are presented in Table 1. Evidently, the most promising decay is $\tau \ar \mu\mu\mu.$ In the Golden Domain we have
\be
2 \cdot 10^{-10} \lesssim {\rm Br}(\tau\ar \mu\mu\mu) \lesssim 6 \cdot 10^{-9},
\ee
which is not too far from the current experimental bound.

A nice feature of the see-saw type II model is that the coupling matrix determines the mass-mixing pattern of neutrinos. In the  Golden Domain the neutrino mass hierarchy  is normal, $\theta_{23}$  lies in the second quadrant, $\theta_{13}$ is moderately large $(2^o-12^o),$ while CP-violating are loosely bounded.

To conclude, we have considered a scenario of lepton flavor violation in the the see-saw type II model which leads to alteration of supernova dynamics and manifests itself in a variety of phenomenological consequences observable in the current and forthcoming experiments, including $\mu\ar e \gamma$ searches at MEG, $\Delta^{\pm\pm}$ searches at LHC, $\tau \ar \mu\mu\mu$ searches at super-B factories,  $\mu-e$ conversion searches at Mu2e (Fermilab) and COMET (J-PARC), $\theta_{13}$ searches in short-base reactor disappearance and accelerator $\nu_e$-appearance experiments. On the other hand, in the considered scenario direct neutrino mass measurement (KATRIN experiment) and $2\beta0\nu$ detection will be unaccessible in the near future due to low neutrino masses.

We have also briefly outlined a scenario of lepton flavor violation in the the singlet bilepton model, which demonstrates drastically different signatures, which, however, also may be probed in the future experiments.

\section*{Acknowledgments}

Our interest to the $\mu\ar e \gamma$ decay was triggered by M.~V. Danilov's talk at the ITEP seminar; we are grateful to him for the interest to our work as well.
We thank S.~I. Blinnikov and V.~A. Novikov for extensive discussions. We acknowledge the partial support from grants NSh-4172.2010.2, RFBR-11-02-00441, RFBR-10-02-01398 and from the Ministry of Education and Science of the Russian Federation under contracts N$^{\underline{\rm o}}$ 02.740.11.5158, 02.740.11.0239. O. L. thanks C. Grojean for useful discussion, I. Antoniadis for hospitality in CERN TH-PH division in a framework of MassTeV project and  M.~E. Shaposhnikov for hospitality in the Laboratory of particle physics and cosmology of EPFL.

\end{document}